\documentclass[
showpacs,preprintnumbers,amsmath,amssymb]{revtex4}
\usepackage{amsmath}

\begin{document}

\tolerance=5000

\def\pp{{\, \mid \hskip -1.5mm =}}
\def\cL{{\cal L}}
\def\be{\begin{equation}}
\def\ee{\end{equation}}
\def\bea{\begin{eqnarray}}
\def\eea{\end{eqnarray}}
\def\beq{\begin{eqnarray}}
\def\eeq{\end{eqnarray}}
\def\tr{{\rm tr}\, }
\def\nn{\nonumber \\}
\def\e{{\rm e}}
\def\ben{\begin{enumerate}}
\def\een{\end{enumerate}}
\def\bei{\begin{itemize}}
\def\eei{\end{itemize}}
\def\ni{\noindent}
\def\bs{\bigskip}
\def\ms{\medskip}


\title{Class of viable modified $f(R)$ gravities describing
inflation and the onset of accelerated expansion}

\author{G. Cognola$^{1}$, E. Elizalde$^{2}$, S. Nojiri$^{3}$,
S.D. Odintsov$^{4}$, L. Sebastiani$^{1}$, and S. Zerbini$^{1}$}

\affiliation{$^{1}$Dipartimento di Fisica, Universit`a di Trento
and Istituto Nazionale di Fisica Nucleare
Gruppo Collegato di Trento, Italia}

\affiliation{$^{2}$Consejo Superior de Investigaciones Cient\'\i
ficas ICE/CSIC-IEEC, Campus UAB, Facultat de Ci\`encies, Torre
C5-Parell-2a pl, E-08193 Bellaterra (Barcelona) Spain}

\affiliation{$^{3}$Department of Physics, Nagoya University, Nagoya
464-8602. Japan}

\affiliation{$^{4}$Instituci\`{o} Catalana de Recerca i Estudis Avan\c{c}ats
(ICREA) and Institut de Ciencies de l'Espai (IEEC-CSIC), Campus UAB,
Facultat de Ci\`encies, Torre C5-Par-2a pl, E-08193 Bellaterra
(Barcelona) Spain}

\date{{\small \today}}


\begin{abstract}

A general approach to viable modified $f(R)$ gravity is developed in both
the Jordan and the Einstein frames. A class of exponential, realistic
modified gravities is introduced and investigated with care. Special
focus is made on step-class models, most promising from the phenomenological
viewpoint and which provide a natural way to classify all viable
modified gravities. One- and two-steps models are explicitly considered,
but the analysis is extensible to $N$-step models.
Both inflation in the early universe and the onset of recent accelerated
expansion arise in these models in a natural, unified way.
Moreover, it is demonstrated that models in this category easily pass all local tests,
including stability of spherical body solution, non-violation of Newton's law,
and generation of a very heavy positive mass for the additional scalar degree of freedom.

\end{abstract}

\pacs{11.25.-w, 95.36.+x, 98.80.-k}

\maketitle

\section{Introduction}

Modified gravity models constitute an interesting dynamical alternative to the
$\Lambda$CDM cosmology in that they are able to describe with success
the current acceleration in the expansion of our Universe, the so called dark energy
epoch. Moreover, modified $F(R)$ gravity (for a review, see e.g. \cite{review})
has undergone many studies which conclude that this gravitational alternative
to dark energy \cite{CDTT,NO} is able to pass
the solar system tests. The investigation of cosmic acceleration
 as well as the study of the cosmological properties of $F(R)$ models has
been carried out in Refs.~\cite{review,CDTT,NO,FR,FR1,cap,lea}.

Recently the importance of those models was reassessed, namely with
the appearance of the so-called `viable' $F(R)$ models
\cite{HS,AB,Uf,UUprd}. Those are models which do satisfy the cosmological
as well as the local gravity constraints, which had caused a number of
problems to some of the first-generation theories of that kind.
The final aim of all these phenomenological models is to describe a segment
as large as possible of the whole history of our universe, as well as to
recover all local predictions of Einstein's gravity, which have been
verified experimentally to very good accuracy, at the solar system scale.

Let us recall that, in general (see e.g. \cite{review}, for a review), the total
action for the modified gravitational models reads
\be
\label{XXX7}
S=\frac{1}{\kappa^2}\int d^4 x \sqrt{-g} \left[R + f(R)\right]+S_{(m)}\, .
\ee
Here $f(R)$ is a suitable function, which defines the modified gravitational
part of the model. The general equation of motion in $F(R)\equiv R+f(R)$
gravity with matter is given by
\be
\label{XXX22}
\frac{1}{2}g_{\mu\nu} F(R) - R_{\mu\nu} F'(R) - g_{\mu\nu} \Box F'(R) +
\nabla_\mu \nabla_\nu F'(R)
= - \frac{\kappa^2}{2}T_{(m)\mu\nu}\ ,
\ee
where $T_{(m)\mu\nu}$ is the matter energy-momentum tensor.

In this paper we investigate two classes of `viable' modified gravitational
models what means, roughly speaking, they have to incorporate the vanishing (or
fast decrease) of the
cosmological constant in the flat ($R\to 0$) limit, and must exhibit a
suitable constant asymptotic behavior for large values of $R$.
A huge family of these models, which we will term
first class ---and to whom almost all of the models proposed in the literature
belong--- can be viewed as containing all possible smooth versions of the
following sharp step-function model. To discuss this toy model, at the
distribution level, will prove to be very useful in order to grasp the
essential features that {\it all} models in this large family are bound to satisfy.
In other words, to extract the general properties of the whole family in
a rather simple fashion (which will involve, of course, precise distribution
calculus).

This simple model reads
\be
f(R)=-2\Lambda_{\rm eff} \, \theta(R-R_0)\,,
\ee
where $\theta(R-R_0)$ is Heaviside's step distribution.
Models in this class are characterized by the existence of one or more
transition scalar curvatures, an example being $R_0$ in the above toy model
(but there can be more, as we will later see).

The other class of modified gravitational models that has been considered
contains a sort of `switching on' of the cosmological constant as a function
of the scalar curvature $R$. A simplest version of this kind reads
\be
f(R)=2\Lambda_{\rm eff}(e^{-bR}-1)\,.
\label{S}
\ee
Here the transition is smooth. The two above models may be combined in a natural
way, if one is also interested in the phenomenological description of the
inflationary epoch. For example, a two-steps model may be the smooth version
of
\be
f(R)=-2\Lambda_0 \, \theta(R-R_0)\,-2\Lambda_{I} \, \theta(R-R_I)\,,
\ee
with $R_0 << R_I$, the latter being the inflation scale curvature.

The typical, smooth behavior of $f(R)$ associated with the one- and two-step models is
given, in the smooth case, in Figs.~\ref{fig1} and \ref{fig2}, respectively.
The main problem associated with these sharp models is the appearance of
possible antigravity regime in a region around the transition point and
antigravity in a past epoch, what is not phenomenologically acceptable. On the other
hand, an analytical study of these models can be easily carried out, as discussed in the
Appendix.

The existence of viable (or ``chameleon'') $f(R)$ theories with a phase
of early-time inflation is not known to us from the literature. The fact
that we are able to provide several classes of models of this kind that
are consistent also with the late-time accelerated expansion is thus a
novelty, worth to be remarked.

\begin{figure}

\begin{center}

\unitlength=1mm
\begin{picture}(100,60)

\put(10,10){\vector(1,0){80}}
\put(10,10){\vector(0,1){40}}

\put(95,10){\makebox(0,0){$R$}}
\put(10,55){\makebox(0,0){$-f(R)$}}

\put(35,5){\makebox(0,0){$R_0$}}
\put(5,41){\makebox(0,0){$2\Lambda_{\rm eff}$}}

\thicklines

\qbezier(10,10)(34,11)(35,25)
\qbezier(35,25)(36,39)(50,40)
\qbezier(50,40)(64,41)(90,41)

\thinlines
\put(35,25){\line(0,-1){15}}
\put(10,41){\line(1,0){80}}

\end{picture}

\end{center}

\caption{\label{fig1} Typical behavior of $f(R)$ in the one-step model).}
\end{figure}
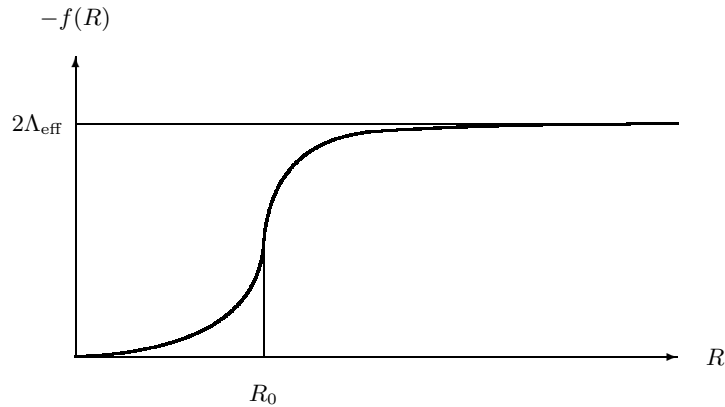

\begin{figure}

\begin{center}

\unitlength=1mm
\begin{picture}(130,85)

\put(10,10){\vector(1,0){110}}
\put(10,10){\vector(0,1){65}}

\put(125,10){\makebox(0,0){$R$}}
\put(10,80){\makebox(0,0){$-f(R)$}}

\put(25,5){\makebox(0,0){$R_0$}}
\put(5,30){\makebox(0,0){$2\Lambda_0$}}
\put(70,5){\makebox(0,0){$R_I$}}
\put(5,70){\makebox(0,0){$2\Lambda_I$}}

\thicklines

\qbezier(10,10)(24.5,10.5)(25,20)
\qbezier(25,20)(25.5,29.5)(40,30)
\qbezier(40,30)(69,31)(70,45)
\qbezier(70,45)(71,69)(120,70)

\thinlines

\put(25,20){\line(0,-1){10}}
\put(40,30){\line(-1,0){30}}
\put(70,45){\line(0,-1){35}}
\put(120,70){\line(-1,0){110}}

\end{picture}

\end{center}

\caption{\label{fig2} Typical behavior of $f(R)$ in the two-step model.}
\end{figure}
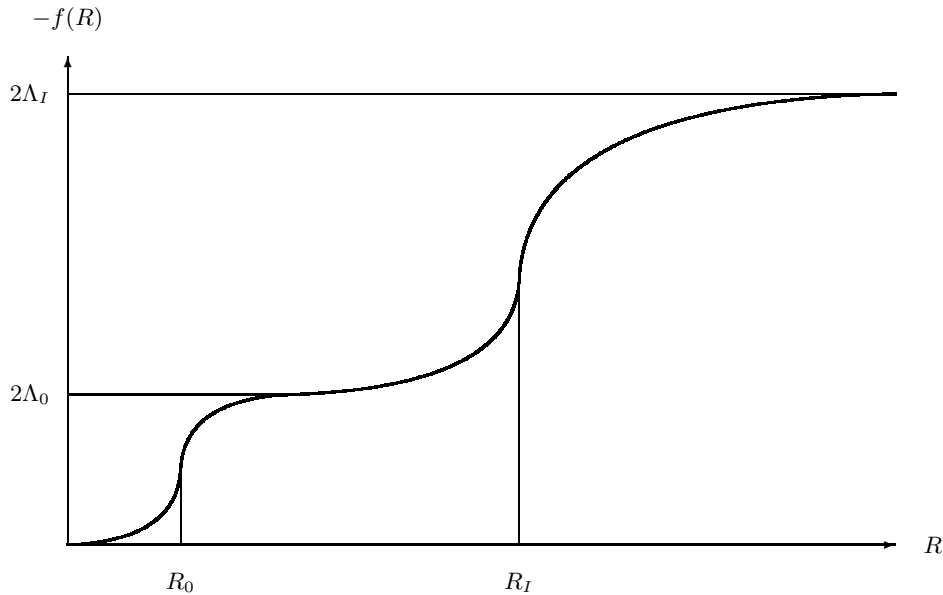

The paper is organized as follows.
Next section reviews $f(R)$ gravity in the physical, Jordan frame.
Equations of motion are presented and the solar system tests (absence of
a tachyon, stability of the spherical body solution, and non-violation of
Newton law) are discussed.
In Sect.~III we give a number of viable modified gravities which may
lead to the unification of early-time inflation with late-time
acceleration and satisfy the solar system tests. Some properties of such
viable modified gravity are discussed in detail.
Sect.~IV is devoted to the presentation of viable models via conformal
transformation, as a kind of scalar-tensor theory which is mathematically
equivalent to the original theory. In some examples, the explicit
form of the scalar potential is derived. Corrections to Newton's law are
also obtained for some of the realistic theories here considered.
It turns out that these models do pass the stringent Newton law bounds, since the
corrections to Newton's law in these cases turn out to be negligible.
Some summary and outlook is given in the Discussions section.
Finally, in an Appendix we show how to evaluate the positions of the corresponding
de Sitter critical points, by considering the sharp version of the one-
and two-step models, expressed in terms of the Heaviside and Dirac distributions.

\section{Modified gravity in the Jordan frame}

Regarding the precise determination of the modifying term, $f(R)$, we here revisit
this issue in the Jordan frame (instead of the Einstein one). Let us recall the two
sufficient conditions which often lead to realistic models (see, for example \cite{HS})
\be
f(0)=0\,, \qquad \lim_{R \rightarrow R_1}f(R)=-\alpha \,,
\label{V}
\ee
where $\alpha$ is a suitable curvature scale which represents an effective
cosmological constant, being $R_1>>R_0$, with $R_0>0$, the transition point.
The condition $f(0)=0$ ensures the disappearance of the cosmological
constant in the limit of flat space-time.

By using these conditions, some models in this class are seen to be able to
pass the local tests (with some extra bounds on the theory parameters)
and are also capable to explain the observed recent acceleration of the universe
expansion, provided that $\alpha=\Lambda_0=2H_0^2$, $H_0$ being the Hubble
constant at the epoch of reference.
However, they do not incorporate early-time inflation, which
comes into play at higher value of $R$. Thus, one might also reasonably
require that \cite{Uf}
\be
f(0)=0\,, \quad \lim_{R \rightarrow R_2}f(R)=-(\alpha+\alpha_I)\,,
\label{VV}
\ee
where $\alpha_I>> \alpha$ is associated with the inflation
cosmological constant, $\Lambda_I$, and where $R_2>>R_I>> R_0$, $R_I$ being the
corresponding  transition large scalar curvature.

Further restrictions, like small corrections to Newton's law and the
stability of planet-like gravitational solutions need to be fulfilled too
\cite{NO}. All those
can be also formulated in the mathematically equivalent Einstein frame. Here
we present a short review of them in the Jordan frame which we consider
as the physical one (see, Ref.~\cite{cap} for a discussion of the physical
(non-)equivalence of the Einstein and Jordan frames).

The starting point is the trace of the equations of motion, which is
trivial in the Einstein theory but gives precious dynamical information in the
modified gravitational models. It reads
\be
3\nabla^2 f'(R)=R+2f(R)-Rf'(R)-\kappa^2 T\,.
\ee
The above trace equation can be interpreted as an equation of motion for
the non trivial `scalaron' $f'(R)$ (since it is indeed associated with the
corresponding scalar field in the other frame). For solutions with constant
scalar curvature $R_*$, the scalaron field is constant
and one obtains the following vacuum solution:
\be
R_*+2f(R_*)-R_*f'(R_*)=0\,.
\label{dS}
\ee

Furthermore, according to \cite{Uf}, we can describe the degree of freedom
associated with the scalaron by means of a scalar field $\chi$, defined by
$F'(R)=1+f'(R)=e^{-\chi}$. If we consider a perturbation around the vacuum
solution of constant curvature $R_*$, given by $R=R_*+\delta R$, where
\be
\label{SS14}
\delta R = - \frac{1+f'(R_*)}{f''(R_*)} \delta\chi \, ,
\ee
then the equation of motion for the scalaron field is
\be
\label{SS16}
\Box \delta\chi - \frac{1}{3}\left(\frac{1+f'(R_*)}{f''(R_*)} - R_*\right)
\delta\chi = - \frac{\kappa^2}{6(1+f'(R_*)}T\, .
\ee
As a result, in connection with the local and with the planetary tests, the
following effective mass plays a very crucial role:
\be
\label{SS19}
M^2\equiv \frac{1}{3}\left(\frac{1+f'(R_*)}{f''(R_*)} - R_*\right)\, .
\ee
If $M^2<0$, a tachyon appears and this leads to an instability.
Even if $M^2>0$, when $M^2$ is small, it is $\delta R\neq 0$ at long ranges,
which generates a large correction to Newton's law. As a result,
$M^2$ has to be positive and very large in order to pass both the local and the
astronomical tests. This stability condition can be also derived within
QFT in de Sitter space-time (see, for instance, \cite{cognola05}).

Concerning the matter instability \cite{DK,NO,Faraoni}, this might occur when
the curvature is
rather large, as on a planet, as compared with the average curvature
of the universe $R\sim \left(10^{-33}\,{\rm eV}\right)^2$.
In order to arrive to a stability condition, we can start by
noting that the scalaron equation can be rewritten in the form
\cite{NO,DK}
\be
\label{XXX23}
\Box R + \frac{f'''(R)}{f''(R)}\nabla_\rho R \nabla^\rho R
+ \frac{(1+f'(R) R}{3 f''(R)} - \frac{2(R+f(R))}{3 f''(R)}
= \frac{\kappa^2}{6f''(R)}T\, .
\ee
If we now consider a perturbation, $\delta R$, of the Einstein gravity
solution $R=R_e=-\frac{k^2T}{2}>0$, we obtain
\be
0\simeq (-\partial_t^2+U(R_e)) \delta R + C\,,
\ee
with the effective potential
\bea
\label{XXX26}
U(R_e)&\equiv& \left(\frac{F''''(R_e)}{F''(R_e)}
 - \frac{F'''(R_e)^2}{F''(R_e)^2}\right)
\nabla_\rho R_e \nabla^\rho R_e + \frac{R_e}{3} \nn
&& - \frac{F'(R_e) F'''(R_e) R_e}{3 F''(R_e)^2} -
\frac{F'(R_e)}{3F''(R_e)}
+ \frac{2 F(R_e) F'''(R_e)}{3 F''(R_e)^2} -
\frac{F'''(R_e) R_e}{3 F''(R_e)^2}\, .
\eea
If $U(R_e)$ is positive, then the perturbation $\delta R$ becomes exponentially
large and the whole system becomes unstable. Thus, the matter stability
condition is, in this case,
\be
U(R_e) <0\,.
\label{MI}
\ee

Coming back to the vacuum condition (\ref{dS}), we recall that,
within the cosmological framework, it may be rederived making use
of the dynamical system approach. This
consists in rewriting the generalized Friedmann equations of the modified
gravitational model in terms of a first-order differential system and looking for
its critical points. For a modified gravity model,
the associated dynamical system can be written as \cite{lea, AG} ($a(t)$
being the expansion factor in a FRW flat space-time):
\bea
\frac{d}{d \ln a}\Omega_{R} &=& 2\Omega_R(2-\Omega_R)\Omega_R)-
\beta (1-\Omega_F-\Omega_\rho)\,, \nn
\frac{d}{d \ln a}\Omega_{F} &=& 2\Omega_F(2-\Omega_R)+(\Omega_F-\Omega_R)
(1-\Omega_F-\Omega_\rho)\,, \nn
\frac{d}{d \ln a}\Omega_{\rho} &=& [2(2-\Omega_R)-3(w+1)+1-\Omega_F-\Omega_\rho]
\Omega_\rho\,,
\eea
where $w=\frac{p}{\rho}$ is the usual barotropic constant, being
\be
\Omega_R=\frac{R}{6H^2}\,,\quad
\Omega_F=-\frac{f(R)-Rf'(R)}{6H^2(1+f'(R))}\,,\quad
\Omega_\rho=\frac{\chi \rho}{3H^2(1+f'(R))}\,,
\ee
with
\be
\beta=\frac{1+f'(R)}{Rf''(R)}\,.
\ee
There exists yet another quantity
\be
\Omega_{\dot F}=-\frac{\dot f'(R)}{H(1+f'(R))}\,,
\ee
which satisfies the constraint
\be
\Omega_{\dot F}+\Omega_{F}+\Omega_{\rho}=1\,.
\ee
The critical points are solutions of the (algebraic) system
\bea
0 &=& 2\Omega_R(2-\Omega_R)\Omega_R)-\beta (1-\Omega_F-\Omega_\rho)\,,\nn
0 &=& 2\Omega_F(2-\Omega_R)+(\Omega_F-\Omega_R)(1-\Omega_F-\Omega_\rho)\,,\nn
0 &=& [2(2-\Omega_R)-3(w+1)+1-\Omega_F-\Omega_\rho]\Omega_\rho\,.
\eea

As an example of those, the de Sitter critical points are the ones in the
invariant
vacuum submanifold $\Omega_\rho=0$. These solutions read
\be
\Omega_R=2\,, \qquad R_*=12H_*^2\,,
\ee
 and
\be
\Omega_F=1\,, \qquad R_*=R_*f'(R_*)-2f(R_*)\,.
\ee
The last equation coincides with Eq.~(\ref{dS}). It is a transcendental
equation,
for all our models, and can be solved only by iteration (see Appendix) or
either
numerically by other methods.

The stability condition associated with the de Sitter critical point
in this dynamical system framework can be investigated too, and reads
\cite{AG}
\be
1<\beta(R_*)=\frac{1+f'(R_*)}{R_*f''(R_*)}\,.
\ee
It coincides with the requirement that the effective mass (\ref{SS19}) be
positive.
In the matter-radiation sector, where $\Omega_\rho$ is non-vanishing, other
critical points may also exist. In the next section, we give explicit
examples of exponential, viable modified gravity.

\section{Examples of realistic exponential modified gravity}

After the general discussion above, we will here present some new viable $f(R)$
models. We start with a most simple one
\be
f(R)=\alpha(e^{-bR}-1)\,.
\label{s}
\ee
Since $f(0)=0$ and $f(R) \rightarrow -\alpha$ for large $R$, conditions
(\ref{V}) are satisfied. Moreover,
\be
f'(R)=-b\alpha e^{-bR}\,,\quad f''(R)=b^2\alpha e^{-bR}\,.
\ee
We have seen that in the discussion of the viability of modified
gravitational models, the existence of vacuum constant curvature solutions
plays a very crucial role, namely the existence of solutions of
Eq.~(\ref{dS}).
With regard to the trivial fixed point $R_*=0$, this model has the
properties
\be
1+f'(0)=1-\alpha b\,, \qquad f''(0)=\alpha b^2\,.
\ee
Thus, the effective mass for $R_*=0$ is
\be
M^2(0)=\frac{1-\alpha b}{3 \alpha b^2}\,,
\ee
and then Minkowski space time is stable as soon as $\alpha b <1$. Such
condition is equivalent to $1+f'(0)>0$.

In order to investigate the existence of other fixed points, we first have to
find the existence conditions and then make use of
Newton's method or some of its variants (see the Appendix). It is easy to see
that for the model (\ref{s}), one has the critical point only if
$ \alpha b >1$, namely $K'(0)>1$, where the function $K(R)$ defined in
Appendix. Then, one can construct an approximation procedure to
solve Eq.~(\ref{dS}), in terms of an iteration process, namely
\be
R_{n+1}=R_{n}-\frac{R_n-R_nf'(R_n)+2f(R_n)}{1+f'(R_n)-R_nf''(R_n)}\,.
\ee
For a starting point, $R=R_1$, large enough, $f(R)$ is approximately
constant and a few iterations give
\be
\quad R_{*,1}\simeq 2\alpha\,.
\ee
This critical point is stable and corresponds to the current acceleration in
the universe expansion. This follows from that fact that the effective mass is
\be
M^2 \simeq \frac{1}{3\alpha b^2} e^{2b \alpha}\, ,
\ee
namely it is positive and large. However, since $ \alpha b >1$,
for very small $R$, one has antigravity effects, namely $1+f'(R)<0$, as
we shall see in Section IV.

Matter instability can also be investigated. Eq.~(\ref{XXX26}) gives
\be
\label{DK1}
U(R_e)\simeq - \frac{1}{3\alpha b^2} e^{2b R_e}\, ,
\ee
which is negative, thus the matter stability condition is fulfilled.

The model we have discussed so far does not exhibit a sharp transition
curvature. But there are many models where one or more transitions of this kind
appear. The Hu-Sawicki (HS) model \cite{HS} belongs to this one-step class
family of models. A simple choice for a one-step model is in our case
\be
f(R)=\alpha \, \left[ \frac{1+e^{-bR_0}}{1+e^{b(R-R_0)}}-1\right]
=-\alpha \frac{e^{bR}-1}{e^{bR}+e^{bR_0}}\,.
\label{A}
\ee
For $R$ very small, we have
\be
f(R)\simeq -\frac{\alpha b}{1+e^{b R_0}}R+O(R^2)\,,
\ee
while for suitable values of $b$, one has the same behavior as in the HS model,
 where the continuous parameter $b$ plays the role of the
integer $n$ in the above mentioned model. Higher values of $b$ give rise to
a sharper transition, occurring at $R_0$ from very small values of $f(R)$
towards a constant value $-\alpha$.
This model has an effective mass, evaluated at $R_*=0$, which turns out to be
negative, thus Minkowski space-time is unstable in this case, and it might
happen that $1+f'(R)< 0$ around the transition.

A simple modification of the above model which incorporates the inflationary era,
namely the requirement (\ref{VV}), is a combination of the two models
discussed so far, that is
\be
f(R)=\alpha(e^{-bR}-1)
-\alpha_I\frac{e^{bR}-1}{e^{bR}+e^{bR_I}}\,,
\label{ABN}
\ee
or, as a two-step model,
\be
f(R)=-\alpha \frac{e^{bR}-1}{e^{bR}+e^{bR_0}}
-\alpha_I\frac{e^{bR}-1}{e^{bR}+e^{bR_I}}\,.
\label{AB}
\ee
Again, $f(0)=0$ and, at the value $R=R_I$, there is a transition to a
higher constant value $-(\alpha+\alpha_I)$ which can be related to inflation.

We should note that $f(R)$ in (\ref{AB}) is a monotonically decreasing function.
Then, when $R$ is very large, $f(R)$ tends to a constant value, which could correspond to the effective cosmological constant generating inflation.
In order to describe the recent accelerating expansion of the universe, $f(R)$ should remain almost constant, that is, $f'(R)=0$, for sufficiently small values of $R$ corresponding to the curvature of the present universe.

The main problem with this models is the appearance of antigravity in
connection with the transition point when the function is sharp (see the
discussion in the Appendix). The appearance of antigravity in the past, namely
around $R_I$, is not acceptable, as we have already commented.

A possible modification of the previous model is the following:
\be
f(R)=-\alpha (e^{-bR}-1)
+cR^N\frac{e^{bR}-1}{e^{bR}+e^{bR_I}}\,,
\label{ABS}
\ee
with $N >2$ and $c>0$. In this variant, similarly to the
theory \cite{UUprd}, during the inflationary era
at $R > R_I$, $f(R)$, the model acquires also a power dependence on the scalar curvature,
which may help to exit from the inflationary stage.

As discussed in detail in the Appendix, for the sharp, theta models, besides the problem of antigravity,
for $R_0 << \alpha$ and $R_I << \alpha_I$, they posses, generically, two De Sitter critical points,
one around the transition point $R_{*}\simeq \frac{5 R_0}{4}$ and the other being
\be
\quad R_{*,2}\simeq 2\alpha\,.
\ee
We can also investigate the matter instability. For the two-step model
(\ref{AB}),
we now assume
\be
\label{DK2}
R_0 \ll R \sim R_e \ll R_I\ .
\ee
Then $f(R)$ in (\ref{AB}) can be approximated as
\be
\label{DK3}
f(R) \sim - \alpha \left\{ -1 + \left(1+\e^{-bR_0}\right)\e^{-b(R-R_0)}\right\}
 - \frac{\alpha_I b R}{1 + \e^{bR_I}}\ .
\ee
We may assume
\be
\label{DK4}
\frac{\alpha_I b}{1 + \e^{bR_I}} \ll 1\ ,
\ee
since $bR_I$ could be very large (see the argument around (\ref{AG5}) about
antigravity). Then we find
\be
\label{DK5}
U(R_e)\simeq - \frac{\e^{b(R_e - R_0)}}{3\alpha b^2\left(1+\e^{-bR_0}\right)}\
,
\ee
which is negative and there is no instability.

We conclude this Section with a variant of the above model which facilitates
 the analytic computation and the discussion concerning antigravity.
In fact, as a smoothed one-step function, we may consider
\be
\label{tan1}
f(R)=-\alpha \left( \tanh \left(\frac{b\left(R-R_0\right)}{2}\right) + \tanh \left(\frac{b R_0}{2}\right)\right)
= - \alpha \left( \frac{\e^{b\left(R-R_0\right)} - 1}{\e^{b\left(R-R_0\right)} + 1}
+ \frac{\e^{b R_0} - 1}{\e^{bR_0} + 1}\right)
\ee
When $R\to 0$, we find that
\be
\label{tan2}
f(R) \to - \frac{\alpha b R}{2\cosh^2 \left(\frac{b R_0}{2}\right) }\ .
\ee
and thus $f(0)=0$, as required. On the other hand, when $R\to +\infty$,
\be
\label{tan3}
f(R) \to - 2\Lambda_{\rm eff} \equiv -\alpha \left( 1 + \tanh \left(\frac{b R_0}{2}\right)\right)\ .
\ee
If $R\gg R_0$ in the present universe, $\Lambda_{\rm eff}$ plays the role of the effective cosmological constant.
We also obtain
\be
\label{tan4}
f'(R)= - \frac{\alpha b }{2\cosh^2 \left(\frac{b \left(R - R_0\right)}{2}\right) }\ ,
\ee
which has a minimum when $R=R_0$:
\be
\label{tan5}
f'(R_0)= - \frac{\alpha b}{2}\, .
\ee
Then in order to avoid antigravity, we find
\be
\label{tan6}
0< 1 + f'(R_0) < 1 - \frac{\alpha b}{2}\, .
\ee
The model given by Eq.~(\ref{tan1}) is able to describe late acceleration. In order to show
that the de Sitter critical points exist, we can compute the function
$K(R)=Rf'(R)-2f(R)$ of the Appendix and we have $K'(R)=Rf''(R)-f'(R)$,
$K''(R)=Rf'''(R)$, $1-K'(0)=1+f'(0)$, where
\be
K'(R)=- \left[bR \tanh \left( \frac{ b(R-R_0)}{2} \right)+1\right]f'(R) \,,
\ee
Thus, enforcing the absence of antigravity, one has
\be
1+f'(0)> 0\,.
\ee
namely, $K'(0)<1$. In this case however $K'(R)> 0$ for $R >R_0$, and it is negative
for $R<R_0$. Therefore, we cannot
use the argument of the Appendix. For the one step-model, however, one can actually
live with antigravity in the future, thus in the sharp version, the analysis
in the Appendix leads again in fact to the existence of two dS critical points.

As a model which is able to describe both the inflation and the late acceleration
epochs, we can consider the following two-step model:
\be
\label{tan7}
f(R)=-\alpha_0 \left( \tanh \left(\frac{b_0\left(R-R_0\right)}{2}\right) + \tanh \left(\frac{b_0 R_0}{2}\right)\right)
 -\alpha_I \left( \tanh \left(\frac{b_I\left(R-R_I\right)}{2}\right) + \tanh \left(\frac{b_I R_I}{2}\right)\right)\ .
\ee
We now assume
\be
\label{tan8}
R_I\gg R_0\ ,\quad \alpha_I \gg \alpha_0\ ,\quad b_I \ll b_0\ ,
\ee
and
\be
\label{tan8b}
b_I R_I \gg 1\ .
\ee
When $R\to 0$ or $R\ll R_0,\, R_I$, $f(R)$ behaves as
\be
\label{tan9}
f(R) \to - \left(\frac{\alpha_0 b_0 }{2\cosh^2 \left(\frac{b_0 R_0}{2}\right) }
+ \frac{\alpha_I b_I }{2\cosh^2 \left(\frac{b_I R_I}{2}\right) }\right)R\ .
\ee
and find $f(0)=0$ again. When $R\gg R_I$, we find
\be
\label{tan10}
f(R) \to - 2\Lambda_I \equiv
 -\alpha_0 \left( 1 + \tanh \left(\frac{b_0 R_0}{2}\right)\right)
 -\alpha_I \left( 1 + \tanh \left(\frac{b_I R_I}{2}\right)\right)
\sim -\alpha_I \left( 1 + \tanh \left(\frac{b_I R_I}{2}\right)\right)\ .
\ee
On the other hand, when $R_0\ll R \ll R_I$, we find
\be
\label{tan11}
f(R) \to -\alpha_0 \left[ 1 + \tanh \left(\frac{b_0 R_0}{2}\right)\right]
 - \frac{\alpha_I b_I R}{2\cosh^2 \left(\frac{b_I R_I}{2}\right) }
\sim -2\Lambda_0 \equiv -\alpha_0 \left[ 1 + \tanh \left(\frac{b_0 R_0}{2}\right)\right] \ .
\ee
Here we have assumed (\ref{tan8b}). We also find
\be
\label{tan12}
f'(R)= - \frac{\alpha_0 b_0 }{2\cosh^2 \left(\frac{b_0 \left(R - R_0\right)}{2}\right) }
- \frac{\alpha_I b_I }{2\cosh^2 \left(\frac{b_I \left(R - R_I\right)}{2}\right) }\ ,
\ee
which has two valleys when $R\sim R_0$ or $R\sim R_I$. When $R= R_0$, we obtain
\be
\label{tan13}
f'(R_0)= - \alpha_0 b_0 - \frac{\alpha_I b_I }{2\cosh^2 \left(\frac{b_I \left(R_0 - R_I\right)}{2}\right) }
> - \alpha_I b_I - \alpha_0 b_0 \ .
\ee
On the other hand, when $R=R_I$, we get
\be
\label{tan14}
f'(R_I)= - \alpha_I b_I - \frac{\alpha_0 b_0 }{2\cosh^2 \left(\frac{b_0 \left(R_0 - R_I\right)}{2}\right) }
> - \alpha_I b_I - \alpha_0 b_0 \ .
\ee
Then, in order to avoid the antigravity period, we find
\be
\label{tan15}
\alpha_I b_I + \alpha_0 b_0 < 2\ .
\ee
The existence of the de Sitter critical points in this two-step model is much
more difficult to investigate. However, in order to get the acceleration of
the Universe expansion it is sufficient that $\omega_{eff} < -\frac{1}{3}$.

We now investigate the correction to the Newton's law and the matter instability issue.
In the solar system domain, on or inside the earth, where $R\gg R_0$, $f(R)$ in (\ref{tan1}) can be approximated by
\be
\label{tan16}
f(R) \sim -2 \Lambda_{\rm eff} + 2\alpha \e^{-b(R-R_0)}\, .
\ee
On the other hand, since $R_0\ll R \ll R_I$, by assuming Eq.~(\ref{tan8b}), $f(R)$ in
(\ref{tan7}) could be also approximated by
\be
\label{tan17}
f(R) \sim -2 \Lambda_0 + 2\alpha \e^{-b_0(R-R_0)}\, ,
\ee
which has the same expression, after having identified $\Lambda_0 = \Lambda_{\rm eff}$ and $b_0=b$.
Then, we may check the case of (\ref{tan16}) only.

We find that the effective mass has the following form
\be
\label{tan18}
M^2 \sim \frac{\e^{b(R-R_0)}}{4\alpha b^2}\, ,
\ee
which could be very large again, as in the last section, and the correction
to Newton's law can be made negligible.
We also find that $U(R_b)$ in (\ref{XXX26}) has the form
\be
\label{tan19}
U(R_e) = - \frac{1}{2\alpha b}\left(2\Lambda + \frac{1}{b}\right)\e^{-b(R_e-R_0)}\ ,
\ee
which could be negative, what would suppress any instability.

Thus, we have here presented several realistic exponential models which naturally
unify the inflation with the dark energy epochs (with a radiation/matter dominance
phase between, as in Refs.~\cite{UUprd,Uf}). In addition, the Newton law is
respected and all spherical body solutions (Earth, Sun, etc) are stable.

\section{Viable models in the Einstein frame}

As is well known from previous studies, it is often quite convenient to
go from the Jordan (physical) frame to the mathematically-equivalent
Einstein frame description, where $f(R)$ models become scalar-tensor
theories with a suitable potential.
In particular, corrections to Newton's law and the matter instability can be
also investigated in the Einstein frame directly, where the relevant
degrees of freedom are a new tensor metric and a scalar field. More specifically,
concerning the inflation issue, the Einstein frame can indeed be very useful.
Following e.g. reference \cite{NO}, we can introduce the auxiliary field $A$
and rewrite the action (\ref{XXX7}) as
\be
\label{XXX10}
S=\frac{1}{\kappa^2}\int d^4 x \sqrt{-g}
\left\{\left(1+f'(A)\right)\left(R-A\right)
+ A + f(A)\right\}\, .
\ee
 From the equation of motion with respect to $A$, if $f''(A) \neq 0$, it
follows that $A=R$.
By using the conformal transformation $g_{\mu\nu}\to \e^\sigma g_{\mu\nu}$,
with $\sigma = -\ln\left( 1 + f'(A)\right)$,
we obtain the Einstein frame action \cite{NO}:
\bea
\label{XXX11}
S_E &=& \frac{1}{\kappa^2}\int d^4 x \sqrt{-g} \left\{ R -
\frac{3}{2}\left(\frac{F''(A)}{F'(A)}\right)^2
g^{\rho\sigma}\partial_\rho A \partial_\sigma A - \frac{A}{F'(A)}
+ \frac{F(A)}{F'(A)^2}\right\} \nn
&&=\frac{1}{\kappa^2}\int d^4 x \sqrt{-g} \left( R -
\frac{3}{2}g^{\rho\sigma}
\partial_\rho \sigma \partial_\sigma \sigma - V(\sigma)\right)\ , \\
V(\sigma) &=& \e^\sigma g\left(\e^{-\sigma}\right) - \e^{2\sigma}
F\left(g\left(\e^{-\sigma}\right)\right)
= \frac{A}{F'(A)} - \frac{F(A)}{F'(A)^2}\, .
\eea
Here $g\left(\e^{-\sigma}\right)$ is given by solving
$\sigma = -\ln\left( 1 + f'(A)\right)=\ln F'(A)$, as
$A=g\left(\e^{-\sigma}\right)$.
After the scale transformation $g_{\mu\nu}\to \e^\sigma g_{\mu\nu}$ is done,
there appears a coupling of the scalar field $\sigma$
with matter. For example, if matter is a scalar field $\Phi$, with mass
$M$, whose action is given by
\be
\label{MN1}
S_\phi=\frac{1}{2}\int d^4x\sqrt{-g}\left(-g^{\mu\nu}\partial_\mu\Phi
\partial_\nu\Phi - M^2 \Phi^2\right)\, ,
\ee
then there appears a coupling with $\sigma$ (in this Einstein frame):
\be
\label{MN2}
S_{\phi\, E}=\frac{1}{2}\int d^4x\sqrt{-g} \left(-\e^{\sigma}
g^{\mu\nu}\partial_\mu\Phi \partial_\nu\Phi
 - M^2 \e^{2\sigma}\Phi^2\right)\, .
\ee
The strength of the coupling is of the same order as that of the gravitational coupling,
$\kappa$. Unless the mass corresponding to $\sigma$, which is defined by
\be
\label{MN3}
m_\sigma^2 \equiv \frac{1}{2}\frac{d^2 V(\sigma)}{d\sigma^2}
=\frac{1}{2}\left\{\frac{A}{F'(A)} - \frac{4F(A)}{\left(F'(A)\right)^2} +
\frac{1}{F''(A)}\right\}\, ,
\ee
is big, there will appear a large correction to the Newton law.
Newton's law has been investigated in the solar system, as well as on earth, where
the curvature is much larger than $R_0$. For the model (\ref{s}), we find
\be
\label{MM1}
m_\sigma^2 \sim \frac{\e^{bR}}{2\alpha b^2}\, ,
\ee
which is positive. As we will discuss soon ((\ref{AG1})-(\ref{AG5})), we find
$1/b\ll R_0 \ll R$,
and therefore $bR\gg 1$, which tell us that $m_\sigma^2$ could be very large and
the correction to the Newton law would be very small. For the model (\ref{AB}), we find
\be
\label{MM2}
m_\sigma^2 \sim \frac{\e^{b(R - R_0)}}{2\alpha b^2\left(1+\e^{-bR_0}\right)}\, ,
\ee
which could be very large again, and the correction to
the Newton law correspondingly very small.

Eq.~(\ref{XXX10}) also tells that if
\be
\label{AG1}
1+f'(A) <0\, ,
\ee
then antigravity could appear, since the effective gravitational constant is
given by
\be
\label{AG2}
\kappa_{\rm eff}^2 \equiv \frac{\kappa^2}{1+f'(A)}\ .
\ee
In order to avoid it, the condition (\ref{AG2}) must be
satisfied, at least until  present, all the way since the beginning of the universe.
Some remark is in order. In the antigravity region, there is no evolution of
the universe with a flat spatial part. In the usual Einstein gravity, we have the
FRW equation, $(3/\kappa^2) H^2 = \rho$, but in the antigravity region, since
the sign of $\kappa^2$ changes, as $\kappa^2 \to - \kappa^2$, we get $-
(3/\kappa^2) H^2 = \rho$. Since the lhs of this equation is always negative
and the rhs is always positive, there is no solution, what shows that
there is no time-evolution of the universe in this case.
We should also note that, even if $1 + f'(A)$ is negative, the conformal
transformation itself can still be well defined, if we use
the absolute value of $1 + f'(A)$, that is, $\left| 1 + f'(A) \right|
g_{\mu\nu} \to g_{\mu\nu}$. In that
case, however, in the obtained Einstein frame action, the sign of the scalar
curvature $R$ becomes negative, that is,  antigravity again appears.
Moreover, the initial value problem which is formulated in $f(R)$ gravity via
conformal transformation \cite{far14} is not well defined. This is the
reason why we avoid the consideration of antigravity regimes.

For the simple model (\ref{S}), the condition (\ref{AG1}) reads
\be
\label{AG3}
1 - \alpha b \e^{-bR}>0\ .
\ee
Since the scalar curvature $R$ in the past universe could be larger than the
curvature $R_0$ in the present universe, we find $1 - \alpha b \e^{-bR} >
1 - \alpha b \e^{-bR_0}$. Therefore, if
\be
\label{AG4}
1 - \alpha b \e^{-bR_0}>0
\ee
is satisfied, the condition (\ref{AG1}) can be satisfied too. Eq.~(\ref{AG4}) tells us that
\be
\label{AG5}
\alpha b<1\ \mbox{or}\ \frac{1}{b}\ll R_0\, .
\ee
In order that $f(R)$ can play the role of an effective cosmological constant
for the present universe, the second condition $1/b \ll R_0$ should be preferred.
The situation is not much changed in the two-step model (\ref{AB}), and the
antigravity condition tells us that $1/b \ll R_0$, again.

We can now investigate the regions which reproduce realistic models.
For the simple model (\ref{S}), conditions (\ref{AG4}) or (\ref{AG5}), which
avoid antigravity in the history of the universe, can be expressed as
\be
\label{AA1}
\frac{1}{b}\ll R_0\sim \left(10^{-33}\,{\rm eV}\right)^2\ .
\ee
On the other hand, the condition that the correction to Newton's law
should be small is that $m_\sigma$, given by (\ref{MM1}), must be large.
Since $\alpha$ in (\ref{MM1}) plays the role of the effective cosmological
constant in the present universe, we have
\be
\label{AA2}
\alpha \sim R_0 \sim \left(10^{-33}\,{\rm eV}\right)^2\ .
\ee
In the solar system, we find $R\sim 10^{-61}\,{\rm eV}^2$.
Even if we choose $1/b \sim R_0 \sim \left(10^{-33}\,{\rm eV}\right)^2$, we
find that
$m_\sigma^2 \sim 10^{1,000}\,{\rm eV}^2$, which is ultimately heavy.
Then, there will not be any appreciable correction to the Newton law.
In the air on earth, $R \sim 10^{-50}\,{\rm eV}^2$, and
even if we choose $1/b \sim R_0 \sim \left(10^{-33}\,{\rm eV}\right)^2$ again,
we find that $m_\sigma^2 \sim 10^{10,000,000,000}\,{\rm eV}^2$. Then, the correction
to Newton's law is never observed in such model.

For the model (\ref{AB}), since $R_0\ll R \ll R_I$ in the solar system or on
the earth, $f(R)$ can be approximated by (\ref{DK3}). Then the effective
gravitational constant could be given by
\be
\label{AA3}
\frac{1}{\kappa^2_{\rm eff}} \equiv \frac{1}{\kappa^2}\left( 1
- \frac{\alpha_I b }{1 + \e^{bR_I}}\right)\, .
\ee
The mass of the scalar field $\sigma$ is given by (\ref{MM2}), which is very
large again, that is, $m_\sigma^2 \sim 10^{1,000}\,{\rm eV}^2$ in the solar system
and $m_\sigma^2 \sim 10^{10,000,000,000}\,{\rm eV}^2$ in the air surrounding the earth,
and therefore the correction to the Newton law is negligibly small, either.

There is a technical point which deserves more careful considerations.
It is that $A=R$ has to be expressed as a function of $\sigma$ by solving the equation
\be
f'(A)=e^{-\sigma}-1
\label{fAsi}\ee
and this can be explicitly done for the simplest cases only.
For example, in Ref.~\cite{HS} a class of models defined by means of the function
\be
f(R)=-\frac{m^2c_1(R/m^2)^n}{1+c_2(R/m^2)^n}\,,\qquad\qquad n\geq1\,.
\label{HS}\ee
has been proposed. Here $c_1,c_2$ are arbitrary dimensionless constants, while
$m$ has the dimension of mass. This model yields an effective cosmological
constant which generates the late-time accelerated expansion.
For such class of models, Eq.~(\ref{fAsi}) reduces to an algebraic equation
of order $2n$, which can be explicitly solved for $n=1$ and $n=2$.
In the simplest case, $n=1$, one easily gets
\be
A_\pm=\frac{m^2}{c_2}\,\left[\pm\,e^{\sigma/2}\sqrt{\frac{c_1}{e^{\sigma}-1}}-1\right]\,,
\qquad\qquad c_2>0\,.
\ee
\be
V(\sigma)=e^{\sigma}\left( 1-e^{\sigma}\right)\,A-e^{2\sigma}f(A)
=\frac{m^2e^{\sigma}}{c_2}\,\left(\sqrt{c_1}\, e^{\sigma /2}-\sqrt{e^{\sigma}-1}\right)^2
\,,\label{}\ee
where the positive solution $A_+$ has been chosen.
For $n=2$ the potential assumes a quite complicated form, which is practically useless.

A simple modification of the model (\ref{HS}) is the following \cite{Uf}:
\be
f(R)=-\frac{m^2c_1(R/m^2)^n+c_3}{1+c_2(R/m^2)^n}\,,\qquad\qquad n\geq1\,,
\ee
which for $n=1$ and $c_2>0$ gives rise to the potential
\be
V(\sigma)=-\frac{m^2e^{\sigma}}{c_2}\,\left[ (c_1+1)e^{\sigma}
-2e^{\sigma /2}\sqrt{(c_1-c_2c_3)(e^{\sigma}-1)}-1\right]\,.
\ee

Now, we go back to the models that we have considered above. For some of them
we can give an explicit form for the potential.
We start with our first model (\ref{V}). In such case, Eq.~(\ref{fAsi})
is a simple transcendental equation which gives rise to
\be
A=-\frac{1}{b}\ln\left[\frac{1}{\alpha b}(1-e^{-\sigma})\right]\,,
\ee
\be
V(\sigma)=\frac{e^{\sigma}}{b}\left[
1+(\alpha b-1)e^{\sigma}+
(e^{\sigma}-1)\,
\ln\frac{1-e^{-\sigma}}{\alpha b}
\right]\,.
\ee
where $\sigma >0$ is understood.

Also for the one-step model (\ref{A}), equation (\ref{fAsi}) becomes
a transcendental one, but it can be solved and we can eventually write
the potential in the form
\be
V(\sigma)=e^{\sigma}\left(1-e^{\sigma}\right)\,A-e^{2\sigma}f(A)\,,
\label{}\ee
where $A=A(\sigma)$ is given by
\be
A_\pm=\frac{1}{\beta}\,\log\left\{
 \frac{\mu e^{\sigma}}{2\gamma(e^\sigma-1)}\,
  \left[
   1-\frac{2(1-e^{-\sigma})}{\mu}\pm\sqrt{1-\frac{4(1-e^{-\sigma})}{\mu}}
  \right]
    \right\}\,,\qquad\qquad \mu=\alpha\beta(1+\gamma)\, ,
\label{}\ee
where $\gamma = e^{-b R_0}$.

Also for the two-step model (\ref{AB}), which includes inflation, one can obtain
an exact expression for the potential $V(\sigma)$ but, since $\alpha_I \gg \alpha$,
such expression reduces to the latter above, with the replacement $\alpha \to \alpha_I$.
The explicit expressions of the scalar potential in the equivalent,
scalar-tensor theory can be actually very useful in the study of the PPN-regime of
modified gravity, in that of the stellar evolution equations, and also in some related
questions.

\section{Discussion and conclusions}

In this paper, a general approach to viable modified gravity  has been
developed in both the Jordan and the Einstein frames. We have focussed
on the so-called step-class models mainly, since they seem to be most promising
from the phenomenological viewpoint and, at the same time, they provide
a natural possibility to classify all viable modified gravities. We have
explicitly presented the cases of one- and two-step models, but a similar
analysis can be extended to the case of an $N$-step model, with $N$ being finite or
countably infinite. No additional problems are expected to appear and the models
can be adjusted, provided one can always find smooth solutions interpolating between
the de Sitter solutions (what seems at this point a reasonable possibility),
to repeat at each stage the same kind of de Sitter transition.
We can thus obtain multi-step models which may lead to multiple inflation and multiple
acceleration, in a way clearly reminiscent of braneworld inflation.

This looks quite promising, with the added bonus that the model's construction
is rather simple, as we have here shown explicitly. All the time,
as a guide for an accurate analysis, use has been made of the simple
but efficient tools provided by the corresponding toy model constructed with
sharp distributions, a new technique that we have here introduced too.
It is to be remarked that, for the infinite-step models, one can naturally expect
to construct the classical gravity analog of the stringy landscape realizations, as
in the classical ideal fluid model \cite{class}.

For the model (\ref{tan7}), both inflation in the early universe and the recent
accelerated expansion could be understood in those models in a unified way.
If we start with large curvature, $f(R)$ becomes almost constant, as in
(\ref{tan10}), and plays the role of the effective cosmological constant,
which would generate inflation. For a successful exit from the inflationary
epoch we may need, in the end, more (say
small non-local or small $R^n$) terms. When curvature becomes smaller, matter
could dominate, what would indeed lower the curvature values.
Then, when the curvature $R$ becomes small enough and $R_0\ll R \ll R_I$,
$f(R)$ becomes again an almost constant function, and plays the role of the
small cosmological constant which generates the accelerated expansion
of the universe, that started in the recent past. Moreover, the model
naturally passes all local tests and
can be considered as a true viable alternative to General Relativity.
Some remark is however in order. On general grounds, one is dealing
here with a highly non-linear system and one should investigate all
possible critical points thereof (including other time-dependent cosmologies),
within the dynamical approach method. Of course, the existence of other
critical points is possible; anyhow, for viable $f(R)$ models, to find them is not
a simple task, and we have here restricted our effort to the investigation of the
dS critical points. With regard to the stability of these points,
the one associated with inflation  should be unstable. In this way, the exit
from inflation could be achieved in a quite natural way. In particular,
for instance, this is in fact the case for the two step model with
the $R^3$ term discussed in the Appendix.

In conclusion, a class of exponential, realistic modified gravities have been here
introduced and investigated with care. Some of these models ultimately lead to the
unification of the inflationary epoch with the late-time accelerating epoch,
under quite simple and rather natural conditions. What remains to be done
is to study those models in further quantitative detail, by
comparing their predictions with the accurate astrophysical data
coming from ongoing and proposed sky observations. It is
expected that this can be done rather soon, having in mind the
possibility to slightly modify the early universe features of the
theories here introduced, while still preserving all of their nice,
realistic current universe properties, as we have shown above.

\vspace{6mm}

\noindent {\bf Acknowledgements.} This paper is an outcome of the
collaboration program INFN (Italy) and DGICYT (Spain). It has been also
supported in part by MEC (Spain), projects FIS2006-02842 and
PIE2007-50/023, by AGAUR (Gene\-ra\-litat de Ca\-ta\-lu\-nya),
grant 2007BE-1003 and contract 2005SGR-00790, by the Ministry of Education,
Science, Sports and Culture of Japan under grant no.18549001 and 21st
Century COE Program of Nagoya University provided by the Japan Society
for the Promotion of Science (15COEG01), and by RFBR, grant 06-01-00609
(Russia).

\appendix

\section{}

In this Appendix, we will study how to evaluate ---or at least get some
information--- on the number and positions of the corresponding
de Sitter critical points. In order to grasp the general behavior of
the larger family of models (as already discussed in Sects. I and II), let us
start by considering the sharp (mathematically very clean albeit physically
unrealistic) version of the one-step models, expressed in terms
of Heaviside and Dirac distributions, namely
\be
f(R)=-\alpha \, \theta(R-R_0)\,, \quad f'(R)=-\alpha \, \delta(R-R_0),
\qquad \alpha >0\,.
\ee
This simple, idealized model leads to antigravity, since $1+f'(R)$ is
obviously always negative at the transition. In practical terms, this means that
the sharper is the smoothing of the step-function, the harder one will be involved
in the antigravity problem. For the one-step model, antigravity could
be arranged to happen in the future. In fact, the equation whose solutions
are the de Sitter critical points becomes
\be
R=-\alpha \, R_0 \, \delta(R-R_0)+2\alpha \, \theta(R-R_0)\,.
\ee
This is an equation in the distribution theory sense and requires
an appropriate treatment. To start, there is the trivial solution $R=0$.
If $R \neq R_0$, the only solution is $R=2\alpha$.
However, eventually we have to deal with a non-ideal, physical situation and
we must consider not the sharp but the smoothed version of the delta and theta
distribution. For the delta, we may consider its support to be contained in
the interval $-\varepsilon+R_0,R_0+\varepsilon$.

We can get information about the other fixed point by arguing as follows.
Integrating the above equation from the value $R=R_0-\varepsilon$ to the value
$R_0+\varepsilon$:
\beq
\frac{1}{2} (4R_0\varepsilon) = - \alpha R_0 + 2\alpha R_0 \varepsilon\,.
\eeq
As  a result, we obtain the consistency condition for $\varepsilon$
\be
\varepsilon \simeq \frac{\alpha}{2(\alpha-R_0)}\,R_0\,.
\ee
Typically, we have  $R_0 << \alpha$. In this case, we obtain
\be
\frac{R_0}{2} < R_{*,1} < \frac{3 R_0}{2}\,\quad
R_{*,2}\simeq 2\alpha\,.
\ee
Thus, with a sharp $f(R)$ function, one has two de Sitter solutions when
$R_0 << \alpha$. antigravity effects can be confined around $R_{*,1}$
(namely, in the future) and the current acceleration is represented by the
second solution $ R_{*,2}\simeq 2\alpha$.

For a two-step sharp model, however, this solution to the problem is
not acceptable, because we cannot allow for antigravity in past
epochs, as discussed in Sect.~IV.
The way out of this is to consider, for example a sufficiently non-sharp
smoothing of the theta functions, but in this case, the above analysis is not longer valid.
Another possibility ---which can still make use of sharp theta functions---
is the following modification of the two-step model, by a power of the curvature:
\be
f(R)=-\alpha \, \theta(R-R_0)+\beta R^N \theta(R-R_I),
\qquad \alpha, \beta >0\,.
\ee
In this case, near the second transition point one does not have, by
construction, antigravity effects. The above analysis can now be applied again,
with the result that, integrating from $R_I$, $N>2$, $N=3$, etc., one gets
\be
R_*\simeq 2\alpha+\beta(N-2) R^N_*\,.
\ee
This is an algebraic equation of N-th order, whose solutions can be
easily investigated, in any specific situation.
For example, for $N=3$ and $2\alpha$ negligible, one  gets the approximate
value
\be
R_*\simeq \frac{1}{\sqrt \beta}.
\ee

In contrast, in the non sharp case it is not so easy
to find sufficient conditions for the existence of the de Sitter critical points.
With regard to this issue, let us recall the following theorem, which may
be useful in a direct, numerical computation of the critical points in a
physically realistic setting. \\
{\sl Theorem}:
Given a twice differentiable function, $G(x)$, defined in the real
interval $[a,b]$ and such that $G'(x)$ is non-vanishing in this interval, and
\be
 |G''(x)G(x)| <|G'(x)^2|\,,
\ee
then the zeroes of $G(x)$ are obtained by the recursive formula
(Newton's tangent method)
\be
x_{n+1}=x_n-\frac{G(x_n)}{G'(x_n)}\,. \label{new1}
\ee
This theorem yields a contraction mapping on the complete metric,
in our case in the interval $[a,b]$. In fact the zeroes of $G(x)$ are the
fixed points of the function $K(x)=x-\frac{G(x)}{G'(x)}$, and this function is
a contraction mapping, as far as $|K'(x)| <1$.
Since $K'(x)=\frac{G''(x)G(x)}{G'^2(x)}$, one gets the stated result.
The recursive relation $x_{n+1}=K(x_n)$ leads to Eq.~(\ref{new1}).

In the cases we consider here, this result is, however, not easy to implement.
Alternatively, one can proceed as follows. Let us write
\be
K(x)=xf'(x)-2f(x)\,,\qquad K(0)=0\,, \ \  K(x)\simeq -2f(x)\,, \ \  x>> 0\,.
\ee
The fixed points $x=K(x)$ are the de Sitter critical points. Let us suppose that
$K'(x)> 0$, for every $x>0$. Then it follows that: \
(i) if $K'(0)>1$ then there exists a fixed point; \
(ii) if $K'(0)<1$ then there are no fixed points.

\end{document}